\documentclass[a4paper,10pt,twocolumn]{article}

\usepackage[english]{babel}
\usepackage[utf8]{inputenc}
\usepackage[T1]{fontenc}
  \usepackage{algorithm2e}

\usepackage[top=1.5cm, left=1.5cm, right=1.5cm, bottom=1.5cm]{geometry}

\renewenvironment{abstract}{\bf\small {\em\ Abstract---}}{}

\usepackage{amsfonts,amssymb,amsmath,amsthm}
\usepackage{subfigure}
\usepackage{graphicx}
\usepackage[footnotesize]{caption}

\usepackage{amsfonts,amssymb,amsmath,amsthm}

\usepackage[pageanchor=false,colorlinks=true,linkcolor=blue,citecolor=blue,urlcolor=blue]{hyperref}
\usepackage{enumitem}

\newcommand*\id{\mathop{}\!\mathrm{d}}

\newcommand\bR{\mathbb{R}}

\newcommand\bC{\mathbb{C}}

\newcommand\sM{\mathcal{M}}

\newcommand\sB{\mathcal{B}}

\newcommand{\tam}{\mathrm{argmin}}

 \newcommand{\ls}{\langle}
 \newcommand{\rs}{\rangle}

\title{An algorithm for non-convex off-the-grid sparse spike estimation with a minimum separation constraint\vspace*{-5mm}}

\author{Yann Traonmilin$^{1,2}$, Jean-Fran\c cois Aujol$^{2}$ and Arthur Leclaire$^{2}$\\
 {\footnotesize $^1$CNRS, $^2$Univ. Bordeaux, Bordeaux INP, CNRS,  IMB, UMR 5251, F-33400 Talence, France.}} \date{\empty} 

\begin{document}
\maketitle
\begin{abstract} Theoretical results show that sparse off-the-grid spikes can be estimated from (possibly compressive) Fourier measurements under a minimum separation assumption. We propose a practical algorithm to minimize the corresponding non-convex functional based on  a projected gradient descent  coupled with an initialization procedure. We give qualitative insights on the theoretical foundations of the algorithm and provide experiments showing its potential for imaging problems.
\end{abstract}

\section{Introduction}

In the space $\sM$ of finite signed measures over $\bR^d$, we aim  at recovering a superposition of impulsive sources $x_0 = \sum_{i=1}^k a_i \delta_{t_i} \in \sM$ from the measurements 
\begin{equation}
 y= Ax_0 + e,
\end{equation}
where $\delta_{t_i}$ is the Dirac measure at position $t_i$, the operator $A$ is a linear observation operator from $\sM$ to $\bC^m$, $y \in \bC^m$ are the $m$ noisy measurements and $e$ is a finite energy observation noise. This inverse problem (called spike super-resolution \cite{Candes_2014,Bhaskar_2013,Tang_2013,Castro_2015,Duval_2015}) models many  problems found in geophysics, microscopy, astronomy or even (compressive) machine learning~\cite{Keriven_2017}. For $l=1,..,m$, 
\begin{equation} \label{eq:distribution}
(Ax)_l =  \int_{\bR^d} \alpha_l(t) \id x(t)
\end{equation}
where $(\alpha_l)_l$ is a collection  of (weighted) Fourier measurements: $\alpha_l(t) =   c_l e^{-j \ls \omega_l,t\rs } $ for some chosen  frequencies $\omega_l \in \bR^d$ and frequency dependent weights $c_l \in \bR$. 
Under a separation assumption on the positions of the Diracs, i.e when $x_0$ is in a set $\Sigma_{k,\epsilon}$ of sums of $k$ $\epsilon$-separated Diracs with bounded support,  it has been shown that $x_0$ can be estimated by solving a non-convex problem  as long as $A$ is an appropriately designed measurement process. This ideal non-convex minimization is: 
\begin{equation} \label{eq:minimization}
    x^* \in \underset{x \in \Sigma_{k,\epsilon}}{\tam} \|Ax-y\|_2^2,
\end{equation}
where 
\begin{equation}
\begin{split}
\Sigma_{k,\epsilon} :=& \left\{ \sum_{r=1}^k a_r \delta_{t_r} :  a \in \bR^k, \right. \\
& \left. \forall  r \neq l, \|t_r-t_l\|_2 \geq \epsilon, t_r \in  \sB_2(R)\right\},\\
\end{split}
\end{equation}
  and $\sB_2(R) = \{t : t_r \in \bR^d, \|t\|_2\leq R \}$ is the $\ell^2$ ball of radius $R$ centered in $0$ in~$\bR^d$. 
 Stable recovery guarantees are obtained for this minimization  when the number of measurements is sufficient. For example, recovery guarantees are obtained when $m \geq O(\frac{1}{\epsilon^d})$ for regular low frequency Fourier measurements~\cite{Candes_2014} and when $m \geq O(k^2d(\log(k))^2 \log(kd/\epsilon) )$ for random Fourier measurements~\cite{Gribonval_2017}. 

First advances in this field proposed a convex relaxation of the problem in the space of measures \cite{Candes_2014,Castro_2015}. While giving theoretical recovery guarantees, these methods are not convex with respect to the parameters due to a polynomial root finding step. They also rely on a SDP relaxation of a dual formulation squaring the size of the problem (which becomes problematic as the dimension $d$ increases). Other methods based on greedy heuristics  (CL-OMP for compressive $k$-means~\cite{Keriven_2017}) have been proposed. Nevertheless, they still lack theoretical justifications in this context while having a good scaling properties with respect to the number of parameters (amplitudes and positions) even if some first theoretical results are emerging for some particular measurement methods~\cite{Elvira_2019}. 

We propose a practical method to solve the non-convex minimization problem~\eqref{eq:minimization} (this abstract is a summary of \cite{Traonmilin_2019b} by the authors). Based on insights from the literature on non-convex optimization for low-dimensional models \cite{Blumensath_2011,Golbabaee_2018,Waldspurger_2018,Ling_2017,Cambareri_2018,Chizat_2019}. Our method relies on two steps : 
\begin{itemize}
 \item  Overparametrized spectral initialization: we propose a spectral initialization step for spike estimation that permits a good first estimation of the positions of the Diracs. 
 \item  Projected gradient descent algorithm in the parameter space:  from \cite{Traonmilin_2018},  the global minimizer of~\eqref{eq:minimization} can be recovered by  gradient descent as long as the initialization lies in an explicit basin of attraction of the global minimizer. It was also shown that projecting on the separation constraint improves the control on the Hessian of the function we minimize. 
\end{itemize}
 
\paragraph{Parametrized formulation}
We consider the following parametrization of $\Sigma_{k,\epsilon}$:  $\sum_{i=1}^k a_i \delta_{t_i} = \phi(\theta)  $ with $\theta= (a_{1},.., a_{k}, t_{1},..,t_{k}) \in \bR^{k(d+1)}$.  We define 
\begin{equation} 
  \Theta_{k,\epsilon}:= \phi^{-1}(\Sigma_{k,\epsilon}),
\end{equation}
the reciprocal image of $\Sigma_{k,\epsilon}$ by $\phi$. Note that any parametrization of elements of $\Sigma_{k,\epsilon}$ is invariant by permutation of the positions. This is not a problem in practice for the convergence of descent algorithms. 
We define the parametrized functional 
\begin{equation}
  g(\theta)  :=  \|A\phi(\theta)-y\|_2^2  
\end{equation}

and consider the problem
\begin{equation} \label{eq:minimization2}
    \theta^* \in \arg \min_{ \theta \in \Theta_{k,\epsilon}}  g(\theta) .
\end{equation}

Since the $\alpha_l$ are smooth, $g$ is a smooth function. Note that performing the minimization~\eqref{eq:minimization2} allows to recover the minima of the ideal minimization~\eqref{eq:minimization}, yielding stable recovery guarantees when $m \geq O(k^2d(\log(k))^2 \log(kd/\epsilon))$ for adequately chosen Gaussian random Fourier measurements and $m \geq O(\frac{1}{\epsilon^d})$ for regular Fourier measurements.   In \cite{Traonmilin_2018}, it has been shown that the simple gradient descent converges  (without projection) to the global minimum of $g$ as long as the initialization falls in an explicit basin of attraction of this global minimum. 
 
\section{Projected gradient descent in the parameter space}

For a user defined initial number of Diracs $k_{in}$, we consider the following iterations: 
\begin{equation}
 \begin{split}
  \theta_{n+1} &= P_{\Theta_{k_{in},\epsilon}}(\theta_n - \tau_n \nabla g(\theta_n))
 \end{split}
\end{equation}
where $ P_{\Theta_{k_{in},\epsilon}}$ is a projection on the separation constraint,
 (notice that there may be several solutions in  $\Theta_{k_{in},\epsilon}$) and $\tau_n$ is the step size at iteration $n$. From~\cite{Traonmilin_2018}, the Hessian of $g$ is better controlled in $\Sigma_{k,\epsilon}$, this prompts us to add the projection step. 
 
The projection  $P_{\Theta_{k_{in},\epsilon}}(\theta)$ could be defined naturally as a solution of the minimization problem 
$\inf_{\tilde \theta \in 
 \Theta_{k_{in},\epsilon}} \|\phi(\tilde \theta) - \phi(\theta)   \|_K$, where $\|\cdot\|_K$ is a metric measuring distance between elements of $\Sigma_{k,\epsilon}$. Unfortunately such optimization is not convex in general.  To avoid this, we  propose a heuristic for $ P_{\Theta_{k_{in},\epsilon}}$ that consists in greedily merging Diracs that are not $\epsilon$-separated:  at each  gradient descent iteration,  merge Diracs that are within balls of radius $\epsilon$ by taking their barycenter.

From \cite{Traonmilin_2018}, this algorithm will converge as soon as the initialization falls into a basin of attraction of a global minimum of $g$. The basins of attraction get larger as the number of measurements increases (up to a fundamental limit depending on the separation $\epsilon$ and the amplitudes in $x_0$). 

\paragraph{Overparametrized spectral  initialization} 

The idea of spectral initialization was used for non-convex optimization in the context of phase recovery~\cite{Waldspurger_2018} and blind deconvolution~\cite{Cambareri_2018}. As we measure the signal $x_0$ at some frequencies $\omega_l$, a natural way to recover an estimation of the signal is to back-project the irregular spectrum on a grid $\Gamma$ that samples $\sB_2(R)$ at a given precision $\epsilon_g$. 
Given a vector of Fourier measurements $y$ at frequencies $(\omega_l)_{l=1,m}$, we calculate 
\begin{equation}
z_{\Gamma,i}=  \sum_{l}y_l  e^{j \ls \omega_l,s_i\rs}
\end{equation}
for some weights $d_l$ to be chosen in the next section (the $s_i \in \Gamma$ are the grid positions). It is possible to show  that when the number of measurements increases and the grid for initialization gets finer, the original positions of Diracs get better localized~\cite{Traonmilin_2019b}. We then perform overparametrized hard thresholding  of the $z_\Gamma$. We propose the initialization $\theta_{init}$ defined by
\begin{equation}\label{eq:initialization}
 \phi(\theta_{init}):=x_{init} = H_{k_{in} }(z_\Gamma)
\end{equation}
where for  $|z_{\Gamma,j_1}| \geq |z_{\Gamma,j_2}| \geq .... |z_{\Gamma,j_n}|$, we have $H_{k_{in}}(z_\Gamma) = \sum_{i=1}^{k_{in}} z_{\Gamma,j_i} \mathbf{1}_{j_i}$.

\section{Numerical Experiments}  \label{sec:exp}

\begin{figure}
 \begin{center}
  \includegraphics[width=0.45\linewidth]{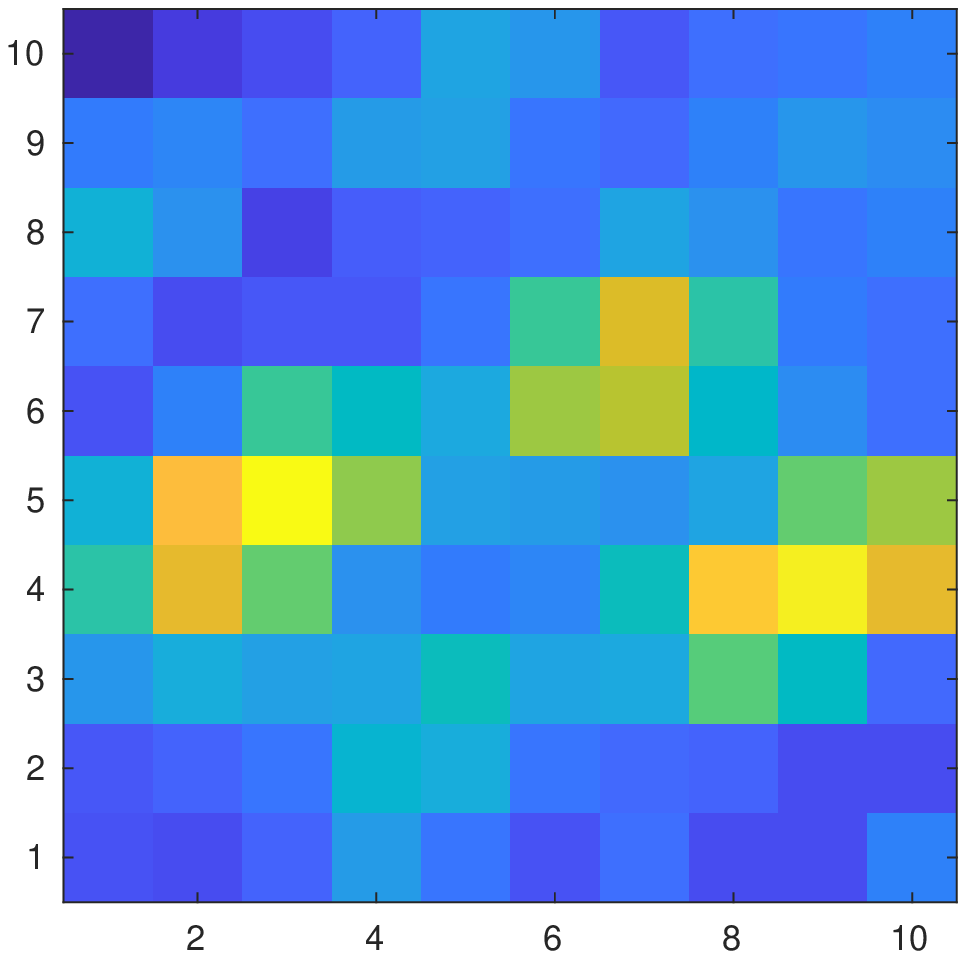} \includegraphics[width=0.45\linewidth]{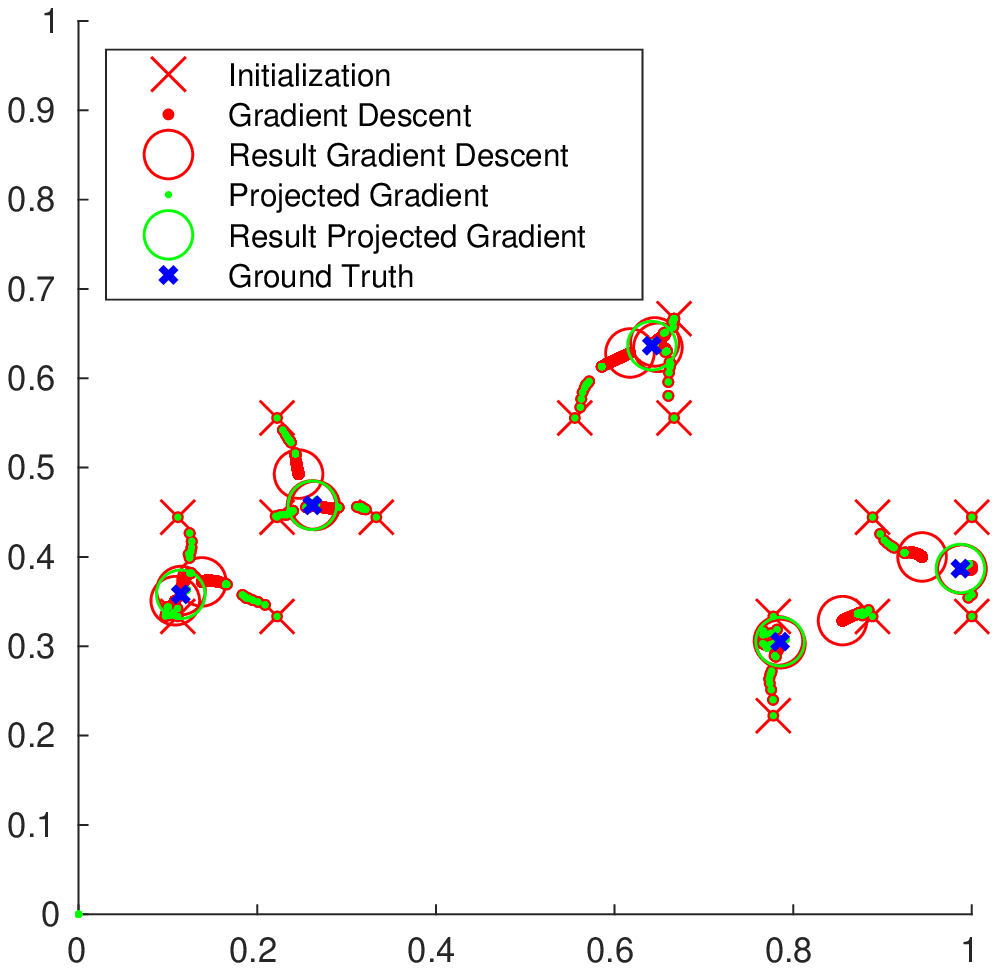}
 \end{center}
\caption{Result for a few spikes in 2d. Left: back-projection of measurements on a grid. Right: Initialization, gradient descent and projected gradient descent trajectories.}\label{fig:proj_grad_simple}\vspace*{-5mm}
\end{figure}

We experiment our algorithm in the noiseless 2d case to show the feasibility of projected gradient descent for imaging applications. The Matlab code used to generate these experiments is available at \cite{Traonmilin_2019code}. As a  proof of concept we perform the recovery of $5$ Diracs in 2 dimensions from $m=120$ Gaussian random measurements. The trajectories of $500$ iterations of the gradient descent and projected gradient descent are represented in Figure~\ref{fig:proj_grad_simple}. We observe that the projection step greatly accelerates the convergence pf gradient descent. 

\begin{figure}[!ht]
 \begin{center}
   \includegraphics[width=0.45\linewidth]{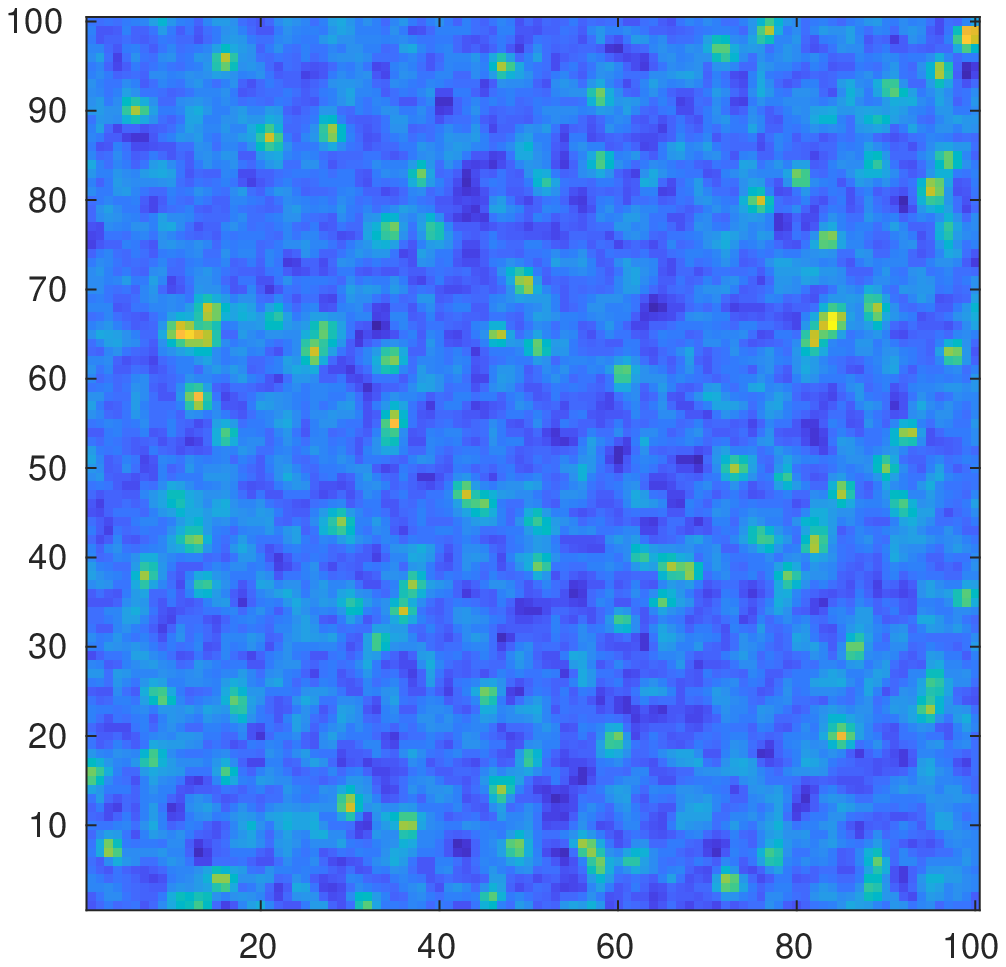}
  \includegraphics[width=0.45\linewidth]{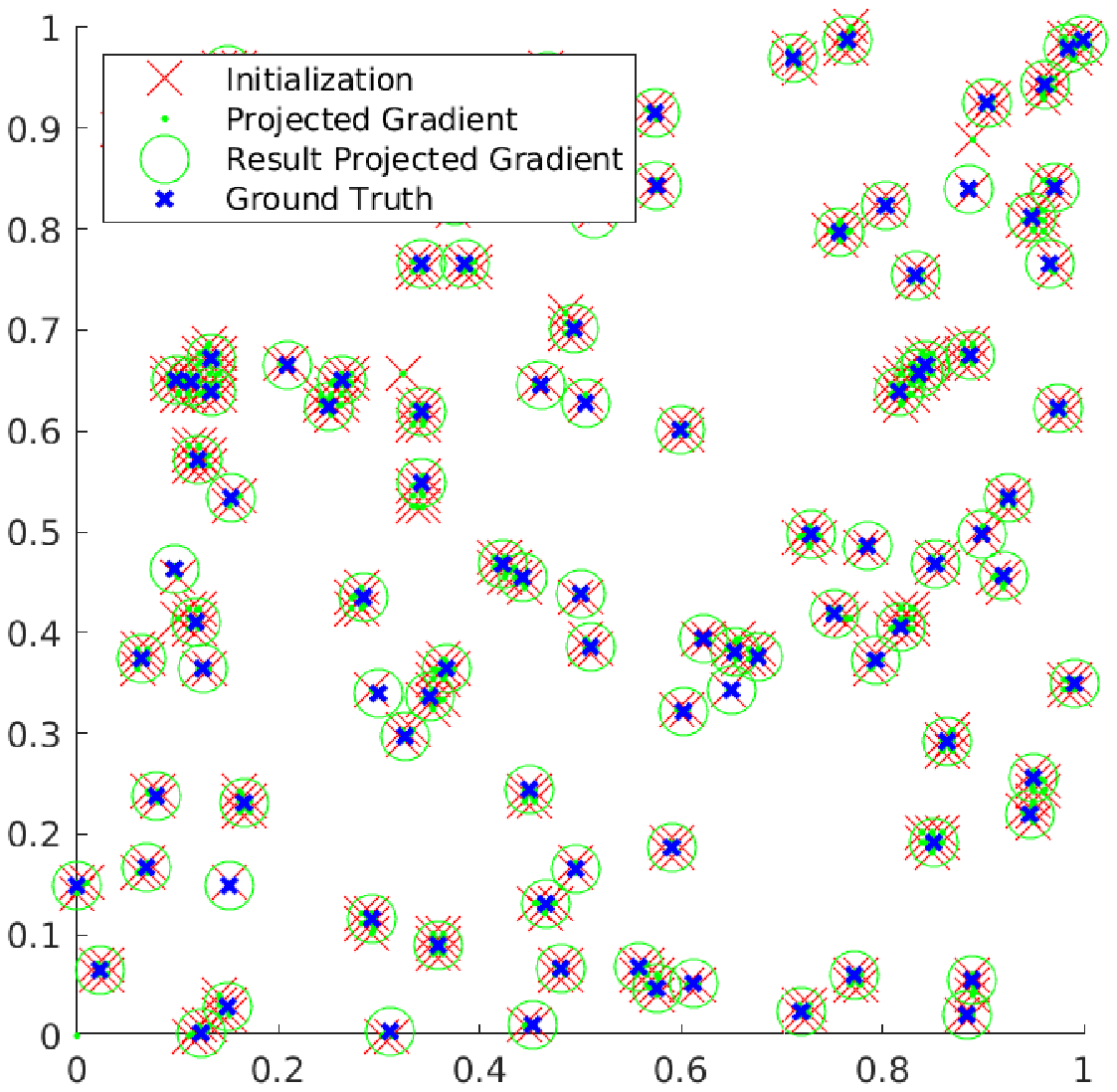}
 \end{center}
\caption{Result for 100 spikes in 2d.  Left: back-projection of measurements on a grid. Right: Initialization, and projected gradient descent trajectories.}\label{fig:proj_grad_big}
\vspace*{-5mm}
\end{figure}

We recover 100 Diracs, with a separation $0.01$ on the square $[0, 1] \times [0, 1]$ from $ m = 2000$ compressive Gaussian measurements (we would need $\approx 10000$ regular measurements to obtain a separation $0.01$). In  practice, the grid $\Gamma$ must be fine enough to overparametrize the number of Diracs with a good sampling of the ideal spectral initialization. If $\epsilon_g$ is too small, the number of initial Diracs needed to sample the energy gets larger, leading to an increased cost in the first iterations   of the gradient descent. In this example we use $\epsilon_g = \epsilon$ and use $k_{in} = 4k$. We observe in Figure~\ref{fig:proj_grad_big} that with these parameters all the Diracs positions are well estimated after $500$ iterations of our algorithm. Similarly as our first example, we  observe that spikes that are not separated in the backprojection  on the grid  (we observe three peaks) are well estimated by our algorithm.

\section{Conclusion} 
We gave a practical algorithm to perform off-the-grid sparse spike estimation in imaging applications, i.e. when the dimension $d$ of the support is not to high (e.g. $d=2$ or $d=3$).  Future research directions are: 
\begin{itemize}
 \item Full theoretical convergence proof of the algorithm with sufficient conditions on the number of measurements. One of the main question is to determine if it is possible to have a convergence guarantee without the computational cost $O((1/\epsilon_g)^d)$ of the backprojection a grid.
 \item A study the algorithm stability to noise and to modeling error with respect to the number of measurements. 
\end{itemize}

\bibliographystyle{abbrv}
\bibliography{proj_grad_itwist2020}

\end{document}